\documentclass[aps,pre,twocolumn,nofootinbib,floatfix]{revtex4-2}

\usepackage{amsmath,amssymb,bm,booktabs,float}
\usepackage{graphicx}
\usepackage{hyperref}

\begin{document}

\title{Learning to Synchronize in Minimum Time}

\author{K. P. O'Keeffe}
\affiliation{Starling Research Institute}

\begin{abstract}
The minimum-time feedback law for driving a population of coupled oscillators into synchrony
is unknown.
Here we settle it for identical Kuramoto oscillators under an instantaneous power constraint.
Greedy control, maximizing $\dot r$ at each instant, is exactly optimal at $N=2$ and
suboptimal above, as dynamic programming confirms at $N=3,4$. The obstruction is geometric:
the greedy closed loop is a reparametrized gradient flow of $r$, fixing its path independently
of the power budget, and as a first-harmonic forcing it cannot leave a single M\"obius orbit.
A three-harmonic policy trained on the DP fields, a trajectory expert, and a smooth
first-hitting-time objective beats greedy by $10$--$14\%$ at
$N=10$--$100$ and matches DP to within $0.3\%$ where ground truth exists. Reading the policy
rather than deploying it collapses it to a two-constant law,
$u_i\propto-\sin\phi_i+a_2\sin2\phi_i+a_3\sin3\phi_i$, which recovers $84$--$99\%$ of the
network's advantage, with the same functional form holding from $N=3$ to $N=1000$.
Machine learning discovered a closed-form law beyond the reach of direct analytical methods.
\end{abstract}

\maketitle

\section{Introduction}
\label{sec:intro}

The minimum-time feedback law for driving a population of coupled oscillators into synchrony is
unknown. Feedback control of collective synchrony was introduced by Rosenblum and
Pikovsky~\cite{rosenblum2004} and given a Lyapunov-based stabilizing design by Sepulchre,
Paley, and Leonard~\cite{sepulchre2007}; neither targets minimum time. For a single oscillator
entrained by an external signal the nearest results maximize the locking range at fixed
power~\cite{harada2010} and minimize the average transient locking time under weak periodic
forcing~\cite{zlotnik2013,snyder2017}; neither is a minimum-time feedback law, and the
collective problem is harder still. Dynamic programming has been applied to small oscillator
populations for minimum energy desynchronization~\cite{nabi2011,nabi2013}; reinforcement
learning has been used to suppress collective activity~\cite{krylov2020}; and the collective
phase of an already-synchronized population has been steered optimally~\cite{fujii2025}. None
of these is a minimum-time feedback law for the order parameter of a synchronizing population.
The statistics of uncontrolled synchronization times have been characterized~\cite{sinha2023},
but the controlled problem is open.

\begin{figure*}[t!]
    \centering
    \includegraphics[width=2\columnwidth]{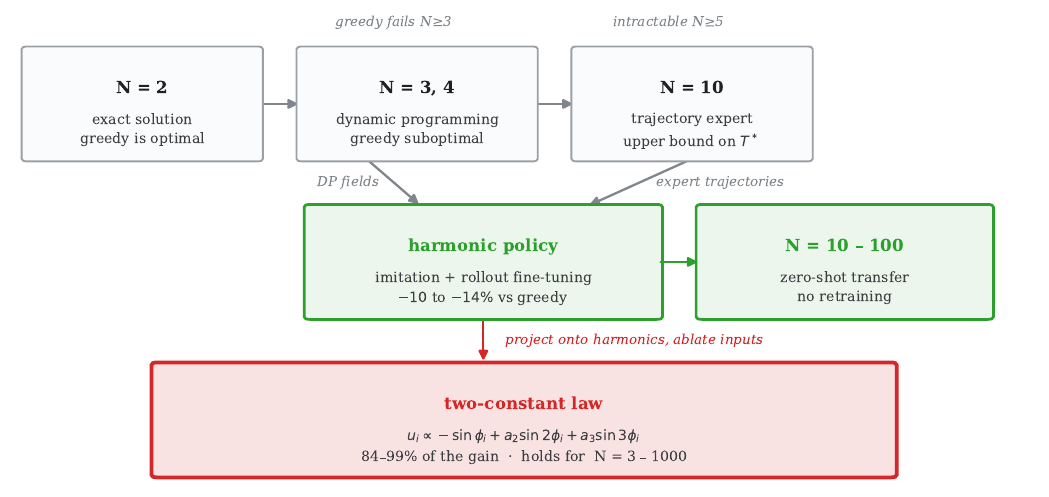}
    \caption{Route of the paper. Top: each exact method and where it runs out --- $N=2$ is
    solvable in closed form, $N=3,4$ require dynamic programming on the reduced state space, and
    the grid becomes intractable for $N\ge5$. Middle: those exact results train a single harmonic
    policy, which transfers to $N=10$--$100$ without retraining. Bottom: projecting that policy
    onto circular harmonics and ablating its inputs returns a two-constant law, which is the
    result.}
    \label{fig:flowchart}
\end{figure*}

Here we answer the question for identical Kuramoto oscillators coupled all-to-all --- a
canonical model in neuroscience, power grids, and beyond~\cite{pikovsky2001,dorfler2014} ---
which synchronize from almost every initial
condition~\cite{kuramoto1984,strogatz2000,acebook2005} but whose transient time depends on the
full phase configuration, not just $r$, the magnitude of the mean-field order parameter~\cite{sinha2023}. Figure~\ref{fig:flowchart} summarizes the route. We
solve the minimum-time HJB equation~\cite{bardi1997} exactly at $N=2,3,4$, establishing that
greedy is suboptimal and providing ground truth for larger-$N$ comparisons. Above $N=4$ the
grid is intractable, so we train a harmonic feedback policy by imitation~\cite{ross2011} on the
DP fields and a trajectory expert at $N=10$. It beats greedy by $10.7$--$14.0\%$ at $N=10,20,50,100$ on
5,000 held-out initial conditions and matches DP to within $0.3\%$ where exact values exist.

The policy is not the result --- it is the probe. Projecting its realized control onto circular
harmonics, ablating its inputs, and scoring every simplification by measured synchronization
time collapses the network to a two-constant formula. That formula transfers in functional form
from $N=3$ to $N=1000$, is nearly invariant across a sixteenfold range of drive power, and
outperforms the network once freed of its three-harmonic architecture~\cite{bronstein2021}.
Learning was the instrument that found the law; the law is the result.

\section{Model and control objective}
\label{sec:model}

We study identical, all-to-all coupled Kuramoto oscillators,
\begin{equation}
    \dot\theta_i = \frac{K}{N}\sum_{j=1}^N \sin(\theta_j-\theta_i) + u_i(t),
    \qquad i=1,\ldots,N,
\end{equation}
with an instantaneous power budget $\sum_i u_i^2\le P$. The order parameter is
$Z=re^{\mathrm{i}\psi}=\frac1N\sum_je^{\mathrm{i}\theta_j}$ and the synchronization time is
$T_{\mathrm{sync}}=\inf\{t\ge0:r(t)\ge r_{\mathrm{target}}\}$. No hold interval is imposed:
under zero control $r=0.95$ already lies inside the basin of the synchronized state, so the
target is absorbing once the drive ceases. Throughout $K=1$, $P=1$, $r_{\mathrm{target}}=0.95$;
after rescaling time the only dimensionless group is $P/K^2$, whose effect we report in
Sec.~\ref{sec:law}.

The control is mean-free: a common offset rotates the frame without changing any phase
difference, so an optimal control can be chosen with $\sum_iu_i=0$, and the minimum-time control
saturates the budget wherever it is defined, so $\|u\|_2=\sqrt P$. Writing $\phi_i=\theta_i-\psi$
for the phases relative to the mean, $\partial r/\partial\theta_i=-\frac1N\sin\phi_i$, so the
\emph{greedy} law maximizing instantaneous $\dot r$ is $u_i^{\mathrm{greedy}}\propto-\sin\phi_i$,
an instance of Fradkov's speed-gradient method~\cite{fradkov2007} and the gradient of the
Lyapunov function used by Sepulchre et al.~\cite{sepulchre2007}.
Note $\nabla r$ is automatically mean-free, since $\sum_i\sin\phi_i=0$ defines $\psi$; the
constraint costs nothing and greedy is the exact pointwise maximizer of $\dot r$.

\section{Exact minimum time, and where it runs out}
\label{sec:hjb}

The minimum time $T^*(\theta_0)$ to reach $r=r_{\mathrm{target}}$ from initial condition
$\theta_0$ satisfies a Hamilton--Jacobi--Bellman equation~\cite{bardi1997}. Working in relative
coordinates $x_k=\theta_{k+1}-\theta_1$, which reduce the $N$-torus by the rotational symmetry,
and writing $B$ for the map from mean-free controls to relative-phase dynamics and $f_0$ for the
uncontrolled drift, the equation is
\begin{equation}
    1 + \nabla T^*\!\cdot f_0 - \sqrt{P}\,\|B^{\mathsf T}\nabla T^*\|_2 = 0,
    \label{eq:hjb}
\end{equation}
with $T^*=0$ on the target. The optimal control
$u^*=-\sqrt P B^{\mathsf T}\nabla T^*/\|B^{\mathsf T}\nabla T^*\|_2$ points opposite to
$B^{\mathsf T}\nabla T^*$: it pushes hardest in the direction $T^*$ falls fastest. Where $T^*$
is not differentiable --- which happens at states with a symmetry, where two equally fast paths
meet and the value function has a kink --- the optimal control is not unique. The worst-case
$N=3$ state below is such a point.

\textbf{Two oscillators.} The reduced state is $d=\theta_2-\theta_1\in[0,\pi]$ with
$\dot d=-K\sin d+v$, $v=u_2-u_1$. Mean-freeness forces $u_2=-u_1$, so the admissible control
set is exactly two points, $v=\pm\sqrt{2P}$: the only choice is which direction to push.
Pushing to close the phase gap is always better --- $v=-\sqrt{2P}$ minimizes $\dot d$
uniformly in $d$ --- so by scalar comparison it reaches the target first. Then
$T^*(d_0)=\int_{d_{\mathrm{tgt}}}^{d_0}\mathrm dd/(K\sin d+\sqrt{2P})$, elementary by the
Weierstrass substitution. Since $r=\cos(d/2)$ is monotone in $d$, $T^*$ is a function of $r$
alone and greedy is exactly optimal. RK4 integration confirms this to better than $0.02\%$.

\begin{figure}[t]
    \centering
    \includegraphics[width=\columnwidth]{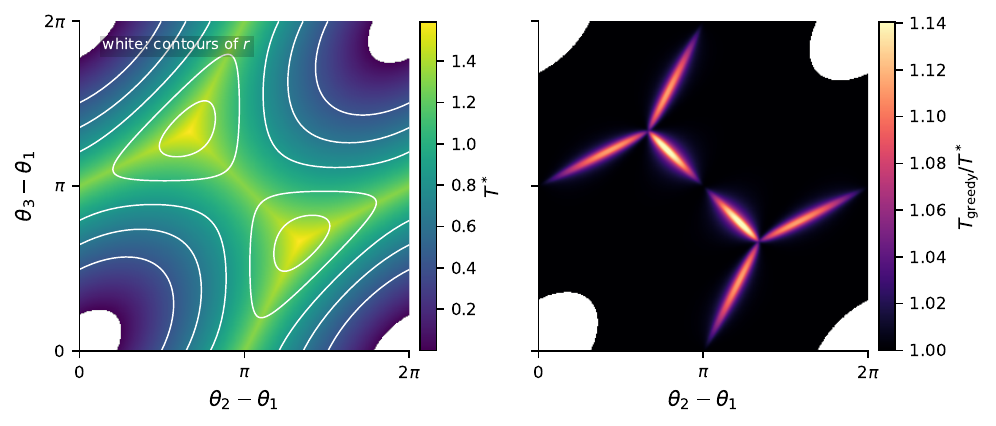}
    \caption{Minimum time at $N=3$ on the reduced torus. Left: $T^*$ with white contours of $r$.
    If greedy were optimal the two families would coincide. Right: the greedy penalty
    $T^{\mathrm{greedy}}/T^*$, concentrated on thin filaments meeting the antipodal critical
    set.}
    \label{fig:n3fields}
\end{figure}

\textbf{Three and four oscillators.} For $N\ge3$ the reduced space is higher dimensional and
$T^*$ need not be a function of $r$. The key diagnostic is whether it is: if $T^*$ depends only
on $r$, greedy is optimal; if $T^*$ varies across states with the same $r$, it is not. We
quantify this via the \emph{level-set spread} --- the range of $T^*$ within a narrow bin of
$r$, divided by its mean, which is zero iff $T^*$ is a function of $r$ alone. We solve
Eq.~(\ref{eq:hjb}) by Lax--Friedrichs sweeping~\cite{kao2004}, cross-checked at $N=3$ against
a semi-Lagrangian scheme~\cite{falcone2013} (Appendix~\ref{app:numerics}). The spread is
$0.224$ at $N=3$ and $0.403$ at $N=4$, both grid converged in the mean (though the worst-case
ratio continues to climb with resolution, as detailed in Appendix~\ref{app:numerics}), and at
least $0.447$ at $N=5$ on a coarser grid where it is still rising.
Figure~\ref{fig:n3fields} shows the $N=3$ field: contours of $T^*$ and of $r$ do not coincide.

A nonzero spread is necessary but not sufficient to prove greedy suboptimal: greedy could still
be optimal at every individual state even if $T^*$ varies across $r$-level sets. The definitive
test is measured time. The DP rollout improves on greedy by $1.0\%$ in the mean at $N=3$ and
$3.5\%$ at $N=4$, with the tail carrying it: p99 improvements of $5.8\%$ and $17\%$. The worst
resolved penalty is $1.148$ at $N=3$ and at least $1.244$ at $N=4$, the latter still growing
with resolution and so a lower estimate.

Above $N=4$ the grid is impractical, and no exact optimal feedback is available. Everything that
follows is built to be checkable against these two value functions.

\section{Learning the controller}
\label{sec:learning}

\textbf{Policy and training.} The policy must respect two symmetries of the problem: rotating
all phases by a common offset leaves the dynamics unchanged, and the oscillators are
exchangeable. We therefore use a rotation-invariant, permutation-equivariant residual policy: an eight-dimensional invariant feature vector built from $r$, $N^{-1/2}$ and pooled
circular moments, a two-layer MLP of width 64, and heads emitting three harmonic amplitudes and
a gate interpolating between greedy and the full harmonic output (5{,}191 parameters,
Appendix~\ref{app:arch}). Mean subtraction and $\ell_2$ normalization enforce the constraints
exactly, and the coefficient head is zero-initialized so an untrained network \emph{is} greedy.
The parameter count does not depend on $N$.

Training has three stages: imitation of the $N=3,4$ value functions, imitation of a
trajectory-optimized expert at $N=10$, and fine-tuning against a smooth first-hitting-time
surrogate. The absorbed hitting time is piecewise constant in the parameters and carries zero
gradient almost everywhere, so the surrogate replaces the indicator by
$q_k=\sigma((r_{\mathrm{target}}-r_k)/\tau_s)$ and the time by
$\widehat T=\delta\sum_k\prod_{j\le k}q_j$, a soft count of pre-crossing steps whose product
acts as an absorbing survival factor. One practical warning, since it cost us a wrong
conclusion: the rollout horizon must contain the crossing. At $N=100$ a horizon of $3$ leaves
$99.8\%$ of initial conditions short of the target, whereupon $q_k\to1$, the surrogate saturates,
$\widehat T$ equals the horizon for every sample, and the time term of the loss has no gradient
at all. Every number we report is a measured crossing of the true dynamics, never a surrogate
value.

\textbf{Performance.} We froze the checkpoint and protocol before opening seeds 20000--24999.
Table~\ref{tab:baselines} gives the result: relative to greedy the policy reduces mean
synchronization time by $10.7$, $14.0$, $13.7$ and $12.0\%$ at $N=10,20,50,100$, with every
trajectory reaching the target and every paired bootstrap interval excluding zero. The same
checkpoint and parameters are used at every $N$ --- no retraining, no fine-tuning per population.
At $N=3,4$ it matches the DP rollout to $0.04\%$ and $0.33\%$,
which is the check that licenses trusting it where DP is unavailable.

The DP fields and expert trajectories are worth one to two percentage points, not the effect.
To verify this: trained from greedy initialization by PPO
against measured time alone --- no value functions, no expert, no imitation --- a policy of the
same architecture reaches within $0.7\%$ of the three-stage pipeline at $N=10$ and $2.2\%$ at
$N=100$. The problem is learnable from scratch.

\section{Reading the controller}
\label{sec:reading}

A 5{,}191-parameter policy that beats a heuristic by 12\% is not by itself informative. We
therefore ask what it computes.

Because the architecture emits harmonic coefficients by construction, projecting the realized
control onto $\{\sin k\phi,\cos k\phi\}$ is exact rather than a fit to a foreign basis --- what
is being extracted is the map from state to coefficients, not a new representation. Two features
appear immediately. The second-harmonic coefficient sits near $+0.5$ throughout the ascent and
the third near $-0.25$, both nearly independent of $r$ and of $N$; and every \emph{cosine}
coefficient averages to zero.

Replacing the network by constants in the same basis costs surprisingly little. Ablating which
collective variables the coefficients are allowed to depend on, and scoring each variant by
measured time on held-out initial conditions, gives the fraction of the network's advantage over
greedy that survives:

\begin{center}
\begin{tabular}{lcc}
coefficients depend on & $N=10$ & $N=100$ \\
\hline
nothing (constants) & 79\% & 94\% \\
$r$ & 81\% & 94\% \\
$r,\,w$ & 79\% & 94\% \\
$r,\,w,\,v$ & 92\% & 97\% \\
$r,\,w,\,v,\,c_3,\,s_3$ & 96\% & 100\% \\
\end{tabular}
\end{center}

\noindent where $w=\langle\cos2\phi\rangle$, $v=\langle\sin2\phi\rangle$,
$c_3=\langle\cos3\phi\rangle$, $s_3=\langle\sin3\phi\rangle$. Neither $r$ nor $w$ contributes
anything; $v$ contributes almost all of the state dependence. That is interpretable: $v$ selects
the \emph{sign} of the symmetry-breaking $\cos\phi$ term, whose mean is zero, which is why a
constant coefficient on it is worthless and a state-dependent one is not. The two-constant law
below forgoes this channel entirely and recovers $84$--$99\%$ regardless; $v$ is what the
remaining few percent are made of. The full affine map
from those five moments to six coefficients --- thirty numbers, fitted at one population size and
applied unchanged at every other --- recovers $97$--$100\%$ of the network's gain at
$N=3,\ldots,100$.

A single second-harmonic term already captures most of the gain. Adding only
\begin{equation}
  u_i \;\propto\; -\sin\phi_i \;+\; c\,\sin 2\phi_i ,
  \label{eq:twoterm}
\end{equation}
with $c\approx0.4$--$0.45$ (close to the $a_2\approx0.5$ read directly from the policy),
recovers $62\%$ ($N=10$) and $65\%$ ($N=100$) of the gain over greedy. Allowing the cheapest
possible state dependence, $c(r)=c_0+c_1 r$ with $c_1<0$, raises this to $68\%$ and $71\%$.
The residual $\sim30\%$ is precisely what a function of $r$ alone cannot supply: it requires
dependence on $w$, which no $c(r)$ can see.

Dropping the cosine terms entirely and adding the third harmonic leaves
\begin{equation}
  \boxed{\;u_i \;\propto\; -\sin\phi_i \;+\; a_2\sin2\phi_i \;+\; a_3\sin3\phi_i\;}
  \label{eq:3term}
\end{equation}
with $a_2\approx0.65$--$0.75$, $a_3\approx-0.25$--$-0.35$. Selecting the two constants on
training seeds and measuring on the frozen block gives the harmonic-law rows of
Table~\ref{tab:baselines} and Fig.~\ref{fig:harmoniclaw}(b): $84\%$ of the network's gain at
$N=10$, rising to $99\%$ at $N=100$. The optimum is flat --- at $N=100$ the whole neighborhood
$a_2\in[0.65,0.85]$, $a_3\in[-0.45,-0.35]$ lands within $0.01$ of the best value --- so the
constants need no tuning to be useful.

The second harmonic cannot simply be increased. With $a_3=0$, every $N=10$ trajectory with
$a_2\ge0.65$ fails to reach the target, locking at $r\approx0.9417$: too strong a second harmonic
stabilizes a two-cluster state. The useful range is bounded by a bifurcation, not by the power
budget, and the third harmonic is what allows $a_2$ past it.

\section{The law is not limited by the architecture that revealed it}
\label{sec:law}

\begin{figure*}[t]
    \centering
    \includegraphics[width=2\columnwidth]{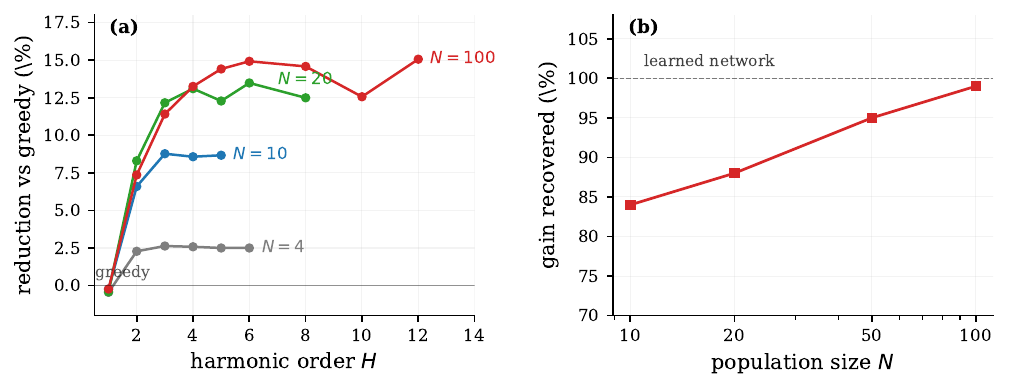}
    \caption{(a) Reduction in mean $T_{\mathrm{sync}}$ relative to greedy against harmonic order
    $H$, coefficients optimized against measured time at each order. $H=1$ buys nothing at any
    $N$; the useful order grows with $N$; gains are exhausted near $H\approx5$. (b) Fraction of
    the network's gain over greedy recovered by the two-constant law, Eq.~(\ref{eq:3term}).}
    \label{fig:harmoniclaw}
\end{figure*}

The network was constrained to three harmonics. The problem is not, and once the law is
written down that ceiling can be lifted. We sweep the harmonic order of
\begin{equation}
  u_i \propto -\sin\phi_i + \sum_{k=2}^{H}\bigl[a_k\sin k\phi_i+b_k\cos k\phi_i\bigr],
  \label{eq:law}
\end{equation}
choosing constant coefficients at each $H$ to minimize measured time on training initial
conditions and reporting on the frozen block. Figure~\ref{fig:harmoniclaw}(a) shows three
things.

\emph{A first-harmonic controller with constant coefficients never beats greedy.} At $H=1$ the
optimizer returns $+0.45$, $+0.24$, $+0.42$ and $+0.23\%$ relative to greedy at
$N=4,10,20,100$ --- all marginally worse, i.e. indistinguishable from it. The exact $N=2$ result
thus extends empirically: greedy is the best constant-coefficient first-harmonic controller at
every population size, and every gain beyond it comes either from higher harmonics or from
letting the first-harmonic coefficients depend on the state. This is a sharper statement than suboptimality, and it says where the missing
time lives. State dependence is worth having, and its value is strongly $N$-dependent. Training $H=1,2,3$
policies through the same pipeline --- whose coefficients, unlike the constant-coefficient sweep
above, vary with the state --- and reporting each against greedy on identical initial conditions
gives

\begin{center}
\begin{tabular}{rcccc}
$N$ & $H=1$ & $H=2$ & $H=3$ & $H=1$ share of $H=3$ \\
\hline
3  & 0.94\% & 0.90\% & 0.98\% & 96\% \\
4  & 2.89\% & 3.08\% & 3.30\% & 88\% \\
10 & 8.28\% & 10.67\% & 10.69\% & 77\% \\
20 & 10.17\% & 13.84\% & 14.70\% & 69\% \\
50 & 6.64\% & 11.28\% & 13.89\% & 48\% \\
100 & 3.05\% & 8.61\% & 12.11\% & 25\% \\
\end{tabular}
\end{center}

\noindent A state-dependent first-harmonic policy is therefore \emph{not} beaten by greedy: it
is WS-preserving yet gains $8$--$10\%$ at $N=10$--$20$. That channel dominates up to
$N\approx20$ and then collapses, and leaving the M\"obius orbit accounts for three quarters of
the gain at $N=100$.
The mechanism behind this transition is explained in Sec.~\ref{sec:mechanism}.

\emph{The useful order grows with $N$.} At $N=100$ the reduction runs $-7.4$, $-11.4$, $-13.3$,
$-14.4$, $-14.9\%$ for $H=2,\ldots,6$. The network's $H=3$ therefore left about three percentage
points on the table at $N=100$: the extracted law, given one more harmonic than its parent
architecture allowed, outperforms the network it was read from (measured on a separate
1,000-initial-condition block, so not directly comparable with Table~\ref{tab:baselines}). Beyond $H\approx5$ the
measurements sit in a band of $-14.4$ to $-15.1\%$ whose spread is dominated by search variance;
the gains are effectively exhausted.

\emph{The law is low-harmonic anyway.} The mean-free control space is $(N-1)$ dimensional and
$H$ harmonics span $2H$ directions, so reaching every admissible control would need $H=50$ at
$N=100$. The gains are exhausted by roughly ten of ninety-nine available directions --- the
structural reason one fixed law works across population sizes.

\emph{Optimized directly, the law matches the network.} Every fit above targeted the network's
controls, so the law could at best tie it. Running CEM on the thirty free affine coefficients
directly against measured $T_{\mathrm{sync}}$ on training seeds and evaluating on the frozen
block at $N=100$ gives $4.093$ versus $4.115$ for the published network ($-0.54\%$,
95\% CI $[-0.70\%,-0.39\%]$, 2,000 paired trajectories). A network retrained with a corrected
horizon reaches $4.070$, $0.55\%$ below the CEM law. The law and a well-trained network
therefore agree to within half a percent; the result worth reporting is the compression, not a
performance win.

\textbf{Robustness.} The constants barely move with the drive power: re-selecting them over a
sixteenfold range of $P$ returns $(0.65,-0.25)$ at $N=10$ for every $P\ge0.5$, with the gain
stable between $8.3$ and $9.1\%$. Held fixed and applied to larger populations, the law beats
greedy by $10.3\%$ at $N=200$, $7.9\%$ at $N=500$ and $5.9\%$ at $N=1000$, every trajectory
reaching. The decay reflects constants selected near $N=50$--$100$, not a failure of form.

\begin{table}
\caption{Synchronization times on the frozen confirmation ensemble (seeds 20000--24999, 5,000
initial conditions per $N$), $K=1$, $P=1$, $r_{\mathrm{target}}=0.95$, RK4 $\Delta t=0.005$. All
rows at a given $N$ use the same initial conditions. ``DP rollout'' replays the feedback
extracted from the value function and is an upper bound on $T^*$, not $T^*$.}
\label{tab:baselines}
\begin{ruledtabular}
\begin{tabular}{llrrrr}
$N$ & policy & mean & median & p90 & p99 \\
\hline
2 & none & 1.292 & 1.105 & 2.836 & 5.290 \\
2 & greedy & 0.442 & 0.410 & 0.975 & 1.180 \\
\hline
3 & greedy & 0.765 & 0.785 & 1.290 & 1.550 \\
3 & DP rollout & 0.758 & 0.785 & 1.265 & 1.460 \\
3 & learned & 0.758 & 0.785 & 1.265 & 1.460 \\
\hline
4 & greedy & 1.016 & 1.035 & 1.555 & 1.990 \\
4 & DP rollout & 0.980 & 1.030 & 1.445 & 1.660 \\
4 & learned & 0.984 & 1.030 & 1.450 & 1.730 \\
\hline
10 & greedy & 1.927 & 1.895 & 2.605 & 3.360 \\
10 & learned & 1.721 & 1.735 & 2.175 & 2.465 \\
10 & harmonic law & 1.749 & 1.760 & 2.230 & 2.530 \\
\hline
20 & greedy & 2.750 & 2.705 & 3.546 & 4.350 \\
20 & learned & 2.364 & 2.373 & 2.845 & 3.145 \\
20 & harmonic law & 2.407 & 2.415 & 2.910 & 3.220 \\
\hline
50 & greedy & 3.863 & 3.850 & 4.605 & 5.320 \\
50 & learned & 3.333 & 3.345 & 3.830 & 4.235 \\
50 & harmonic law & 3.357 & 3.360 & 3.875 & 4.290 \\
\hline
100 & greedy & 4.675 & 4.675 & 5.370 & 5.930 \\
100 & learned & 4.115 & 4.115 & 4.635 & 5.025 \\
100 & harmonic law & 4.119 & 4.120 & 4.650 & 5.040 \\
\end{tabular}
\end{ruledtabular}
\end{table}

\section{Why greedy gives up time}
\label{sec:mechanism}

The law's structure has a clean explanation: two exact identities say why greedy gives up time
and where it goes.

\textbf{Greedy is a gradient flow of $r$.} The mean-field drift is
$Kr\sin(\psi-\theta_i)=KNr(\nabla r)_i$ and greedy is $\sqrt P\nabla r/\|\nabla r\|_2$, so the
greedy closed loop is
\begin{equation}
  \dot\theta=\Bigl[\,KNr+\sqrt P/\|\nabla r\|_2\Bigr]\nabla r ,
  \label{eq:gradflow}
\end{equation}
exactly, for every $N$, $K$, $P$ (verified to $4\times10^{-16}$). Greedy trajectories are the
gradient \emph{lines} of $r$: fixed curves, independent of $K$ and $P$, with only the traversal
speed set by the budget. Greedy cannot re-route when given more power. It stalls exactly at the
critical set of $r$ --- the antipodal-cluster states, where $\nabla r$ and the drift vanish
together --- and it is time-optimal iff the $r$-gradient line is the time-optimal path,
automatic at $N=2$ and generically false above. This predicts an insensitivity we observe: over
a fortyfold range of $P$ the p99 penalty moves from $1.069$ to $1.057$ at $N=3$ and $1.104$ to
$1.109$ at $N=4$.

\textbf{What a first-harmonic drive cannot steer.} With
$w=\mathrm{Re}(e^{-2\mathrm{i}\psi}\frac1N\sum_je^{2\mathrm{i}\theta_j})$ and
$\sum_i\sin^2\phi_i=\frac N2(1-w)$,
\begin{equation}
  \dot r=\frac{Kr}{2}(1-w)+\nabla r\cdot u,
  \qquad \max_{\|u\|_2\le\sqrt P}\nabla r\cdot u=\sqrt{\frac{P(1-w)}{2N}},
  \label{eq:rw}
\end{equation}
exactly and at finite $N$. Both the drift term and the ceiling on the control term improve as
$w$ falls, so $w$ is worth lowering twice over --- and the only harmonic that drives it directly
is the second. This is Eq.~(\ref{eq:3term})'s $a_2$ term, and the measurement confirms the
mechanism: the learned policy sits below greedy in $w$ at every matched $r$, by $0.13$ at
$N=10$, while giving up about $5\%$ of the instantaneous ascent rate to do so and recovering
more than twice that in total time.

Globally, identical oscillators under common first-harmonic forcing are Watanabe--Strogatz
integrable~\cite{watanabe1993,watanabe1994}, evolving by a M\"obius action with $N-3$ constants
of motion~\cite{marvel2009,pikovsky2015,chen2017,ott2008,ott2009}. Greedy has exactly this form, so it is
confined to one M\"obius orbit --- a corollary of Ref.~\cite{marvel2009}, whose hypothesis
already admits state-dependent forcing that is common across oscillators; the new part is
reading the foliation as a reachability obstruction. Mixing harmonic orders breaks it; a
\emph{pure} higher harmonic remains reducible~\cite{gong2019,skardal2011,vlasov2016}. That the optimal
control actually leaves the orbit is measurable at $N=4$: the M\"obius cross-ratio drifts by
$8.6\times10^{-13}$ uncontrolled and $2.3\times10^{-11}$ under greedy --- both at the integration
floor --- against $1.6\times10^{-1}$ under DP-optimal control, on $99.4\%$ of initial conditions.
At $N=3$ there are no constants, and there Eq.~(\ref{eq:gradflow}) is the whole story.

\section{How close to optimal?}
\label{sec:optimality}

Neither the network nor the law is proven optimal, so we bracket the remainder.
Appendix~\ref{app:lb} derives $\dot r\le G_N(r)=Kr(1-r^2)+\sqrt{P(1-r^2)/N}$ for every
admissible control, hence a lower bound by quadrature; a per-initial-condition trajectory
optimizer supplies an achieved upper value. At $N=100$ on held-out initial conditions these give
$2.630\le T^*\le3.990$ against a measured $4.134$, so between $3.5$ and $36\%$ of the time
remains recoverable. The two sides are far apart because the bound is loose, not the controller.
At $N=3,4$, where $T^*$ is known exactly, essentially the entire gap to the bound is bound slack.
The optimizer demonstrates at least $3.5\%$ remains recoverable; that gap grows with $N$ and
points to the same cause as the DP-only ablation: a law found at small $N$ is not quite the
right law at large $N$.

\section{Discussion}
\label{sec:discussion}

Greedy ascent of the order parameter is minimum-time only for two oscillators, and for larger
populations it is the best first-harmonic controller with constant coefficients. The missing
time lives in state dependence and in harmonics greedy cannot express, and a law with two constants recovers most of it --- one
that transfers from three oscillators to a thousand, is nearly invariant across a sixteenfold
range of drive power, and outperforms the network from which it was read once given a fourth
harmonic.

The route matters as much as the destination. Direct solution stops at $N=4$; a trained policy
does not, but a policy is not an explanation. Reading the policy --- in a basis its own
architecture already used, ablating its inputs, and scoring every simplification by measured
performance rather than by fidelity to the network --- turned a checkpoint into a formula, and
the formula then outran the checkpoint. The DP ground truth was essential for this: it provided
the verification that licensed trusting the policy where exact values are unavailable, and the
initial imitation signal that shaped what the network learned. A practitioner who only wants the
$10$--$14\%$ gain can get it with PPO alone; the DP fields are what made the gain interpretable.

Two limitations bound the scope. The oscillators are identical, which is what makes the
Watanabe--Strogatz structure exact; with a frequency spread the law's advantage over greedy
survives essentially undiminished to a standard deviation of $0.2$, but at $0.4$, where the
target becomes marginally reachable, the strong second harmonic becomes a liability and greedy
reaches more often. And the controller accelerates a transient whose endpoint is already
guaranteed. Below the locking threshold, where control must change the attractor rather than the
approach to it, whether a comparably compact law exists is open. A natural extension is to
swarmalators~\cite{okeeffe2017,okeeffe2022}, where oscillators also move in space and
minimum-time synchronization acquires a spatial component. The values of the two constants
are not yet explained analytically; a perturbative calculation near the incoherent state is the
natural place to look.

\section*{Data Availability}

Code, frozen protocols, checkpoints, raw paired synchronization times, and scripts regenerating
every table and figure are deposited at \url{https://github.com/kevinkawchak/Tsync_RL}.

\appendix

\section{Policy architecture and training}
\label{app:arch}

Input features are $r$, $N^{-1/2}$ and pooled circular moments $\langle\cos k\phi\rangle$,
$\langle\sin k\phi\rangle$ up to $k=4$; a two-layer MLP of width 64 with $\tanh$ activations; a
coefficient head giving $2H=6$ harmonic amplitudes and a scalar gate head whose sigmoid
interpolates between greedy and the harmonic output. Total 5{,}191 parameters, independent of
$N$. Both heads are zero-initialized, so the untrained policy is exactly greedy. The
per-oscillator control is mean-subtracted and $\ell_2$-normalized, enforcing $\sum_iu_i=0$ and
$\|u\|_2=\sqrt P$ to machine precision.

The fine-tuning objective is $\mathbb E[\widehat T]$ plus a worst-decile tail term, a terminal
softplus deficit and a cosine imitation term, with $T_H=3$, $\delta=0.02$, $\tau_s=0.02$, batch
32, learning rate $10^{-4}$, 500 steps at $N=10$. Seeds 0--799 are used for training and
800--999 for checkpoint selection; the confirmation block 20000--24999 was opened only after the
checkpoint and protocol were frozen.

\section{Numerical schemes for the HJB equation}
\label{app:numerics}

Equation~(\ref{eq:hjb}) is solved on the $(N-1)$-torus by Lax--Friedrichs
sweeping~\cite{kao2004} with central differences and viscosity coefficients bounding
$|\partial H/\partial p_k|$ uniformly. The greedy solve uses the \emph{same} coefficients as the
optimal solve; the natural smaller bound for the greedy Hamiltonian gives unequal numerical
diffusion and spurious violations of $T^{\mathrm{greedy}}\ge T^*$. Updates are Jacobi on the
GPU. At $N=3$, Lax--Friedrichs and semi-Lagrangian agree to $0.6\%$.

One caveat governs use of these fields. Ratio statistics converge long before the absolute value
function: refining $N=4$ from $M=256$ to $M=512$ moves the mean spread under $1\%$ and $\max T^*$
by $0.05\%$, while the worst-case \emph{ratio} still climbs, $1.134\to1.244$, because an extremum
over a grid is dominated by the few cells where $T^*$ is smallest and least resolved. The
absolute $T^*$ also carries an upward viscosity bias large enough that a grid value must never be
compared against a simulated crossing. We quote ratios and percentiles from the grids and
measured times everywhere else.

\section{Lower bound on the minimum synchronization time}
\label{app:lb}

With $\Sigma=\sum_i\sin^2\phi_i$, and since $\nabla r$ is mean-free with
$\|\nabla r\|_2=\sqrt\Sigma/N$,
\begin{equation}
  \dot r = \frac{Kr}{N}\Sigma + \nabla r\cdot u,
  \qquad \max_{\|u\|_2\le\sqrt P}\nabla r\cdot u=\frac{\sqrt{P\Sigma}}{N}.
\end{equation}
Maximizing over configurations at fixed $r$ requires maximizing $\Sigma=N-\sum_i\cos^2\phi_i$
subject to $\sum_i\cos\phi_i=Nr$; Cauchy--Schwarz gives $\Sigma\le N(1-r^2)$, attained for even
$N$ by the balanced two-cluster state and unattainable for odd $N$. Hence
$\dot r\le G_N(r):=Kr(1-r^2)+\sqrt{P(1-r^2)/N}$, and by the comparison lemma --- which requires
no monotonicity of $r$, so a trajectory that first \emph{lowers} $r$ only falls further below the
comparison solution ---
\begin{equation}
  T^*(\theta_0) \ge T_{\mathrm{lb}}(r_0)
  = \int_{r_0}^{r_{\mathrm{target}}} \frac{\mathrm dr}{G_N(r)} .
  \label{eq:lb}
\end{equation}
At $N=2$ the whole state space is the extremal family, so the bound is exact; the measured ratio
is $1.004$, the residual being time-step quantization. At $N=3,4$ it holds on all 5,000 initial
conditions with zero violations. Resolving the odd-$N$ slack numerically at $N=3$ raises
$T_{\mathrm{lb}}$ from $0.667$ to $0.715$.

The control term scales as $\sqrt{P/N}$, so at large $N$ the bound is drift limited,
$T_{\mathrm{lb}}\simeq(2K)^{-1}\ln N$ for random initial conditions with $r_0\sim N^{-1/2}$,
while measured times grow as $\approx K^{-1}\ln N$. The factor of two is identifiable: the bound
saturates $\Sigma=N(1-r^2)$, the balanced two-cluster value, whereas random configurations sit
near $\Sigma\approx N/2$. The bound cannot certify optimality at large $N$; it certifies that no
controller improves on the measured times by more than $23\%$ at $N=10$ or $36\%$ at $N=100$.

\bibliography{refs}

\end{document}